\documentclass[useAMS,usenatbib]{mn2e}

\usepackage{hyperref}
\usepackage{graphicx}
\usepackage{amssymb}

\hypersetup{pdftitle={The power spectrum from the angular distribution of galaxies in the CFHTLS-Wide fields at redshift 0.7}
  pdfauthor={Benjamin R. Granett},
  hyperfootnotes=true,
  colorlinks=true,
  pdfborder={0 0 0},
  citecolor=black,
  urlcolor=blue,
  filecolor=blue,
  linkcolor=black,
}

\newcommand{\simlt}{\,\hbox{\lower0.6ex\hbox{$\sim$}\llap{\raise0.6ex\hbox{$<$}}}\,}

\newcommand{\LCDM}{$\Lambda$CDM}

\newcommand{\etal}{et al.~}

\newcommand{\Cl}{\mathcal{C}_l}
\newcommand{\Pl}{\mathcal{P}_l}

\newcommand{\n}{{\bf \hat{n}}}

\newcommand{\sqrdeg}{${\rm deg}^2$}

\newcommand{\z}{{\rm z}}
\newcommand{\zbar}{\bar{{\rm z}}}
\newcommand{\zphot}{{\rm z_{phot}}}
\newcommand{\hmpc}{{\rm ~h~Mpc^{-1}}}
\newcommand{\nside}{{\rm n_{side}}}
\newcommand{\dd}{\partial}
\newcommand{\dlambda}{\partial\lambda}
\newcommand{\dC}{\partial{\bf C}}
\newcommand{\Cinv}{{\bf C}^{-1}}
\newcommand{\C}{{\bf C}}
\newcommand{\Tr}{{\rm Tr}}
\newcommand{\dCdi}{\frac{\dd\C}{\dlambda_i}}
\newcommand{\dCdj}{\frac{\dd\C}{\dlambda_j}}
\newcommand{\dCdk}{\frac{\dd\C}{\dlambda_k}}
\newcommand{\x}{{\bf x}}
\newcommand{\lam}{{\bf \lambda}}

\title[The galaxy power spectrum at z$\sim$0.7]{The power spectrum from the angular distribution of galaxies in the CFHTLS-Wide fields at redshift $\sim$0.7}

\author[Granett et al.]{B. R. Granett$^{1}$\thanks{E-mail:\href{mailto:ben.granett@brera.inaf.it}{ben.granett@brera.inaf.it}},
  L. Guzzo$^{1}$,
  J. Coupon$^{2}$,
  S. Arnouts$^{3}$,
  P. Hudelot$^{4}$,
  O. Ilbert$^{5}$,
\newauthor
  H. J. McCracken$^{4}$,
  Y. Mellier$^{4}$,
  C. Adami$^{5}$,
  J. Bel$^{6}$,
  M. Bolzonella$^{7}$,
  D. Bottini$^{8}$,
\newauthor
  A. Cappi$^{7}$,
  O. Cucciati$^{9}$,
  S. de la Torre$^{10}$,
  P. Franzetti$^{8}$,
  A. Fritz$^{8}$,
  B. Garilli$^{8,5}$,
\newauthor
  A. Iovino$^{1}$,
  J. Krywult$^{11}$,
  V. Le Brun$^{5}$,
  O. Le Fevre$^{5}$,
  D. Maccagni$^{8}$,
  K. Malek$^{12}$,
\newauthor
  F. Marulli$^{7,13,14}$,
  B. Meneux$^{15}$,
  L. Paioro$^{8}$,
  M. Polletta$^{8}$,
  A. Pollo$^{12,16,17}$,
  M. Scodeggio$^{8}$,
\newauthor
  H. Schlagenhaufer$^{18,15}$,
  L. Tasca$^{5}$,
  R. Tojeiro$^{19}$,
  D. Vergani$^{20,7}$~and
  A. Zanichelli$^{21}$
\\
    $^{1}$ Istituto Nazionale di Astrofisica - Osservatorio Astronomico di Brera, Via Brera 28, 20122 Milano, Via E. Bianchi 46, 23807 Merate, Italy\\
    $^{2}$ Astronomical Institute, Graduate School of Science, Tohoku University,  Sendai 980-8578, Japan\\
    $^{3}$ Canada-France-Hawaii Telescope, 65--1238 Mamalahoa Highway, Kamuela, HI 96743, USA\\
    $^{4}$ Institute d'astrophysic de Paris, UMR7095 CNRS, Universit\`{e} Pierre et Marie Curie, 98 bis Boulevard Arago, 75014 Paris, France\\
    $^{5}$ Laboratoire d'Astrophysique de Marseille (UMR 6110), CNRS-Universit\'e de Provence, 38, rue Fr\'ed\'eric Joliot-Curie, \\\hspace{1em} 13388 Marseille Cedex 13, France\\
    $^{6}$ Centre de Physique Th\'eorique, UMR 6207 CNRS-Universit\'e de Provence, Case 907, F-13288 Marseille, France\\
    $^{7}$ Istituto Nazionale di Astrofisica - Osservatorio Astronomico di Bologna, via Ranzani 1, I-40127, Bologna, Italy\\
    $^{8}$ Istituto Nazionale di Astrofisica - Istituto di Astrofisica Spaziale e Fisica Cosmica Milano, via Bassini 15, 20133 Milano, Italy\\
    $^{9}$ Istituto Nazionale di Astrofisica - Osservatorio Astronomico di Trieste, via G. B. Tiepolo 11, 34143 Trieste, Italy\\
    $^{10}$ SUPA, Institute for Astronomy, University of Edinburgh, Royal Observatory, Blackford Hill, Edinburgh EH9 3HJ, UK\\
    $^{11}$ Institute of Physics, Jan Kochanowski University, ul. Swietokrzyska 15, 25-406 Kielce, Poland \\
    $^{12}$ Center for Theoretical Physics of the Polish Academy of Sciences, Al. Lotnikow 32/46, 02-668 Warsaw, Poland\\
    $^{13}$ Dipartimento di Astronomia, Alma Mater Studiorum - Universit\`a di Bologna, via Ranzani 1, I-40127 Bologna, Italy\\
    $^{14}$ INFN/National Institute for Nuclear Physics, Sezione di Bologna, viale Berti Pichat 6/2, I-40127 Bologna, Italy\\
    $^{15}$  Max-Planck-Institut f\"{u}r Extraterrestrische Physik, D-84571 Garching b. M\"{u}nchen, Germany\\
    $^{16}$ Astronomical Observatory of the Jagiellonian University, Orla 171, 30-001 Cracow, Poland\\
    $^{17}$ The Andrzej Soltan Institute for Nuclear Studies, ul. Hoza 69, 00-681 Warszawa, Poland \\
    $^{18}$ Universit\"{a}tssternwarte M\"{u}nchen, Ludwig-Maximillians Universit\"{a}t, Scheinerstr. 1, D-81679 M\"{u}nchen, Germany\\
    $^{19}$ Institute of Cosmology and Gravitation, Dennis Sciama Building, University of Portsmouth, Burnaby Road, Portsmouth, PO1 3FX\\
    $^{20}$  Istituto Nazionale di Astrofisica Istituto di Astrofsica Spaziale e Fisica Cosmica Bologna, via Gobetti 101,I-40129 Bologna, Italy\\
    $^{21}$ Istituto di Radioastronomia Istituto Nazionale di Astrofisica, via Gobetti 101, I-40129, Bologna, Italy\\
}

\begin{document}

\date{}

\pagerange{\pageref{firstpage}--\pageref{lastpage}} \pubyear{2011}

\maketitle
\label{firstpage}

\begin{abstract}
We measure the real-space galaxy power spectrum on large scales at
redshifts 0.5 to 1.2 using optical colour-selected samples from the
CFHT Legacy Survey.  With the redshift distributions measured with a
preliminary $\sim$14000 spectroscopic redshifts from the VIMOS Public
Extragalactic Redshift Survey (VIPERS), we deproject the angular
distribution and directly estimate the three-dimensional power
spectrum. We use a maximum likelihood estimator that is optimal for a
Gaussian random field giving well-defined window functions and error
estimates.  This measurement presents an initial look at the
large-scale structure field probed by the VIPERS survey.  We measure
the galaxy bias of the VIPERS-like sample to be $b_g=1.38 \pm 0.05$
($\sigma_8=0.8$) on scales $k<0.2\hmpc$ averaged over $0.5<{\rm
  z}<1.2$.  We further investigate three photometric redshift slices,
and marginalising over the bias factors while keeping other \LCDM~
parameters fixed, we find the matter density $\Omega_m=0.30\pm0.06$.

\end{abstract}

\section{Introduction}
The shape of the galaxy clustering power spectrum encodes the
dynamical history of the Universe under the influence of baryons, dark
matter and dark energy.  On large scales the assumption of Gaussianity
can be made, and the statistic summarises all of the cosmological
information that is available.  

Measurements of the power spectrum at $\z\sim0$ have led to
fundamental tests of the \LCDM~model \citep{Tegmark04b,Efstathiou02}.
The angular distribution of galaxies on the sky, although less
sensitive than the full three-dimensional view, has played an
important role as well.  Indeed, strong tests of the CDM model were
made with the two-dimensional correlation function from the APM galaxy
survey \citep{Maddox90}.  Photometric surveys have the capability to
probe significantly larger volumes at higher sampling rates than
targeted spectroscopic surveys.  The advantages have become clear with
the advancement of photometric redshift estimation methods.  The loss
in three-dimensional precision can be compensated for with increased
statistics leading to strong cosmological constraints that are
comparable to the results from spectroscopic surveys.

Additionally, the projected density field is only weakly sensitive to
redshift-space distortions, thus it provides a means to infer the
real-space power spectrum directly.  The dependence on peculiar
velocities becomes important for narrow redshift slices and can be
turned into a useful measure of the growth rate \citep{Ross11}.
Measurements of the baryon acoustic feature and redshift-space
distortions have now been made on photometric samples taken from the
Sloan Digital Sky Survey \citep{Padmanabhan07,Blake07,Thomas11}.

In this analysis, we present a new measurement of the real-space
galaxy power spectrum using a photometric catalogue of galaxies at
$0.5<\rm{z}<1.2$ from the Canada-France-Hawaii Telescope Legacy Survey
(CFHTLS) Wide survey.  The survey consists of four fields covering a
total area of 133 \sqrdeg.  The extent of the largest field, W1, is
$\sim10\degr$ or $200~{\rm h}^{-1}~{\rm Mpc}$ at z=0.7, giving a
maximum scale we may probe of $k_{min}\sim0.05\hmpc$. The dataset has
been used for previous cosmological analyses, in particular for weak
lensing \citep{Fu08,Kilbinger09,Tereno09,Shan11} and galaxy
correlation function measurements \citep{Coupon11}.

A key ingredient needed to interpret the projected density field and
constrain the three-dimensional power spectrum is the redshift
distribution of the galaxy sample.  For this we use spectroscopy from
the VIMOS Public Extragalactic Survey\footnote{VIPERS website:
  \href{http://vipers.inaf.it}{vipers.inaf.it}}
(VIPERS)\citep{VIPERS}.  VIPERS is an ongoing spectroscopic program to
target $10^5$ galaxies in the redshift range 0.5-1.2 in a total area
of 24\sqrdeg~in the CFHTLS W1 and W4 fields. The accuracy of the
spectroscopic measurements from VIPERS provide an unbiased estimate of
the redshift distribution.  With this knowledge, we are confident that
we can deproject the angular clustering signal and constrain the three
dimensional power spectrum.  The primary advantage of studying the
deprojected power spectrum $P_k$ is its closeness to theory.  The
shape of the angular power spectrum is complicated by its dependence
on survey properties, its depth and geometry.  Furthermore, in
projection, scales are mixed.  Ideally we would like to separate the
power on large scales in the linear regime from power on small scales
that is influenced by complex astrophysical processes.

\begin{figure}
\includegraphics[scale=1]{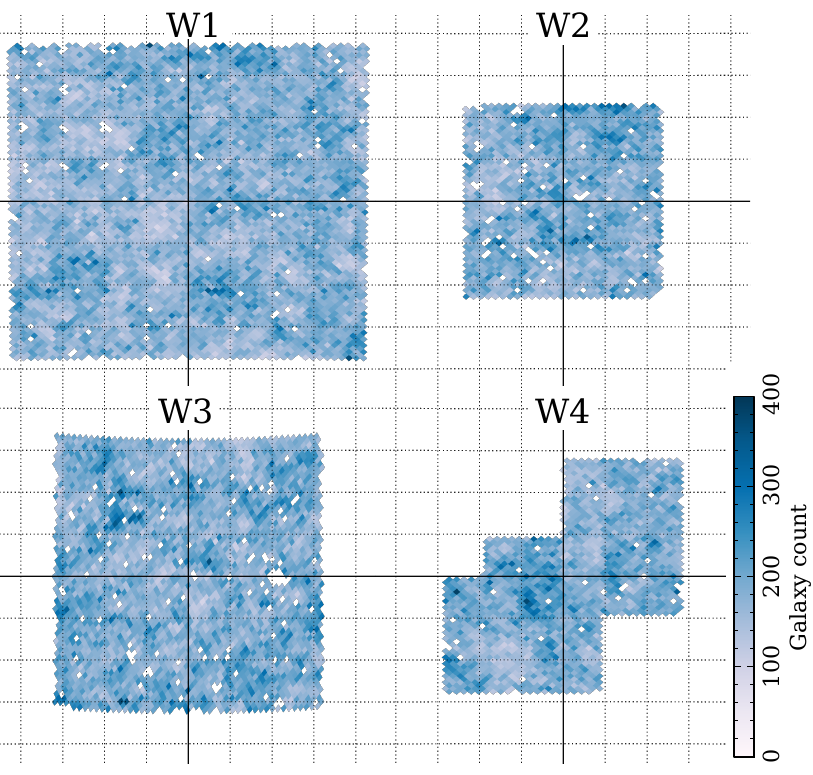}
\caption{Galaxy count maps in the CFHTLS-Wide fields with the
  VIPERS-like colour selection.  We use {\sc Healpix} cells with size
  7\arcmin.  Gaps in the survey coverage are left as blank pixels.
  The grid overlay has spacing of 1\degr.\label{fig:fields}}
\end{figure}

How to derive the three-dimensional power spectrum from the
two-dimensional density field is a problem of deconvolution.  A good
inversion method should be stable against noise in the data.
Preliminary work done by \citet{Baugh93,Baugh94} and
\citet{Gaztanaga98} used the Lucy deconvolution method that is known
to be robust.  To further derive cosmological constraints, we must be
able to estimate the covariance of the deprojection.  Methods of
propagating the error from the angular correlation function to the
three-dimensional power spectrum were developed by \citet{Dodelson00}
who perform the inversion with a prior on the smoothness of the power
spectrum and compute a covariance matrix of the estimate.  Further
work by \citet{Eisenstein01}, \citet{Dodelson02} and \citet{Maller05}
made use of the singular-value decomposition technique to remove modes
that destabilise the inversion.

Importantly, the deprojection method should produce well-defined
window functions that describe the mode mixing.  The aim is to
separate the small and large scales that are mixed in projection, and
the residual leakage should be understood.  A similar problem was
solved with the maximum likelihood methods developed for the cosmic
microwave background angular power spectrum \citep{Tegmark97,Bond98}
and then later applied to galaxy surveys \citep{Huterer01,Tegmark02}.
Applications to the deprojection of the power spectrum were presented
by \citet{Efstathiou01} and \citet{Szalay03}.

In this work, we adopt the maximum-likelihood technique to construct
an estimator for the power spectrum. The result is optimal under the
assumption that the density is represented by a Gaussian random field.
This is a reasonable assumption for the galaxy distribution on large
scales.  Moreover, the estimator also simultaneously gives the
covariance of the estimate as well as the window functions.  For small
surveys, where the window functions must be handled carefully, the
approach is especially useful.  We pay close attention to the window
functions for the results presented here.  Maximum likelihood
estimates are computationally expensive. However, because of the
relatively small field sizes we consider, we can perform all
computations on a consumer level four-core desktop computer.

In this article, we first introduce the CFHTLS and VIPERS datasets
used.  In Sections 3 and 4 we review the angular power spectrum
formalism and the maximum likelihood deprojection using a quadratic
estimator.  We then apply the method to Gaussian simulations and
investigate potential biases due to uncertainty in the redshift
distribution and the fiducial cosmology.  Lastly, we measure
the power spectrum with CFHTLS data and constrain the linear galaxy
bias and matter density.

We report magnitudes using the AB magnitude convention in the CFHT
$u^*g'r'i'z'$ photometric system.  We assume a flat \LCDM~cosmology with
$H_0=70.4~{\rm km~s^{-1}~Mpc^{-1}}$, $\Omega_m=0.272$,
$\Omega_b=0.0456$, $n_s=0.963$ and $\sigma_8=0.8$ \citep{Larson11}.

\section{Data}
\subsection{Photometric selection}
The Canada-France-Hawaii Telescope Legacy Survey (CFHTLS) Wide
includes four fields labelled W1, W2, W3 and W4.  The total area is
133\sqrdeg~imaged with five-band photometry $ugriz$ to a depth of
$i=24.5$.  We construct colour-selected galaxy samples from these
fields to match the spectroscopic target selection used by the VIMOS
Public Extragalactic Survey (VIPERS).

VIPERS is a spectroscopic program to measure the redshifts of galaxies
over an area of 24\sqrdeg~in the CFHTLS W1 and W4 fields.  Galaxies
are targeted from CFHTLS-Wide photometry to a flux limit of
$i_{AB}=22.5$ with colour criteria to produce a sample at ${\rm
  z}>0.5$ having few low-redshift interlopers.  The selection is done
in the $u-g$, $r-i$ colour plane with the following limits on
extinction-corrected magnitudes: (1) $r-i \ge 0.7$ AND $u-g \ge 1.4$
OR (2) $r-i\ge 0.5(u-g)$ AND $u-g < 1.4$ \citep{VIPERS}.  We replicate
these colour criteria on the full CFHTLS-Wide photometry T0006 release
\citep{CFHTLS}.  Hereafter, we refer to this selection as the
VIPERS-like sample.

Every source has an estimated photometric redshift and star-galaxy
classification from the T0006 photometric redshift catalogue.  The
star-galaxy classification accounts for both the source profile and
fits to stellar spectral templates \citep{Coupon09}.  We apply the
same criteria as used for the VIPERS target selection and exclude all
sources photometrically classified as stars (7\% of sources). From the
remaining sample identified as galaxies, we remove sources that fail
to fit galaxy spectral templates with reduced $\chi^2>100$.  This cut
removes sources with incomplete or spurious photometry amounting to
$\sim$0.15\% of sources which are typically near to the edges of
masked regions, bright stars or field borders.

We also use the photometric redshifts to select sub-samples of
narrower slices in redshift, labelled S6 ($0.5<{\rm z_{phot}}<0.6$),
S7 ($0.6<{\rm z_{phot}}<0.8$) and S8 ($0.8<{\rm z_{phot}}<1.$), see
Table \ref{table:samples}.  As with the VIPERS-like sample, these are
also limited to $i_{AB}=22.5$.  We have confirmed that these
photometric redshift selections at $z>0.5$ also meet the VIPERS
selection criteria.  Thus, the VIPERS spectroscopy can be used to
calibrate the redshift distributions of these samples without
introducing a bias.

The VIPERS spectroscopic targets were selected from the CFHTLS T0005
catalogue after field-to-field colour corrections were applied by the
VIPERS team.  These colour corrections are no longer necessary in the
T0006 update, and it has been shown that the selections from the two
catalogue versions match well \citep{VIPERS}.  We note that a limited
area in the CFHTLS has been observed with a replacement $i$-band
filter called $y$.  The photometric redshifts were computed with the
appropriate filter transmission function. To construct our $i<22.5$
limited samples, we take a reasonable approach and do not distinguish
between the two bands.

The CFHTLS catalogues also include corrections for Galactic
extinction.  We do not consider residual systematic effects of
reddening here because all four fields have relatively low and uniform
extinction at the level of $E(B-V)=0.06$ in W4 and $<0.02$ in the
other fields.  The $i=22.5$ limit is 2 magnitudes brighter than the
detection limit of the survey, thus we do not expect the selection to
be affected by the extinction correction.

\begin{figure}
a)\includegraphics[scale=1]{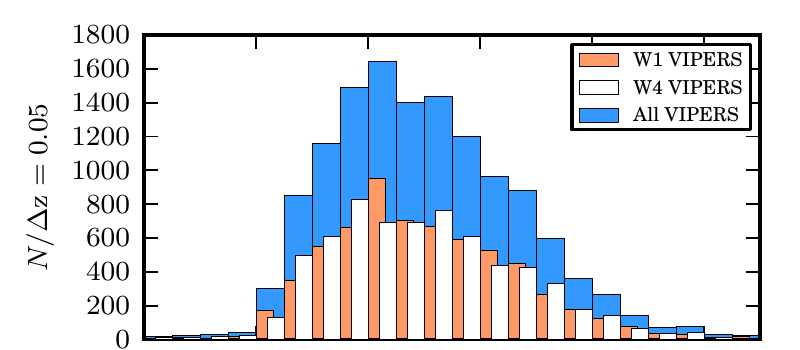}
b)\includegraphics[scale=1]{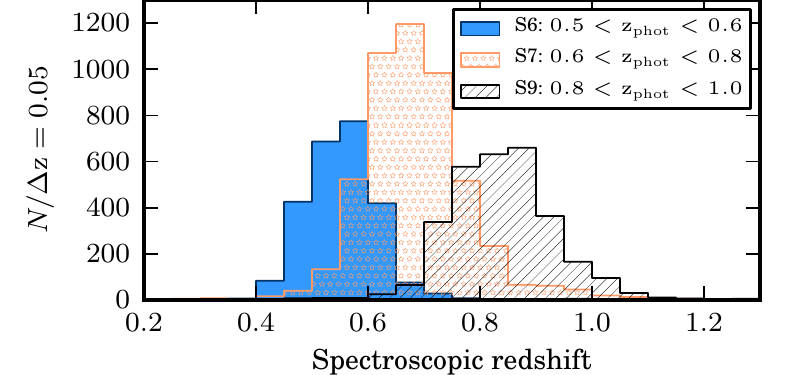}
\caption{Redshift distributions of the spectroscopic samples used.
  (a) The distribution of the full VIPERS sample is shown.
  Overplotted are the distributions from the W1 and W4 fields
  individually. (b) The spectroscopic redshift distributions of the
  three photometric redshift subsamples are plotted.\label{fig:zdist}}
\end{figure}

\subsection{Density maps}
We construct the density maps for the photometrically-selected samples
by counting galaxies in cells defined by the {\sc Healpix} scheme with a
resolution of 7\arcmin~($\nside=512$) \citep{Healpix}.  The maps for the
VIPERS-like colour selection are shown in Fig. \ref{fig:fields}.  We
use a survey mask provided by CFHTLS and exclude sources that fall
within the halos of bright stars.  For cells that fall on a mask
boundary, we measure the fractional coverage using a uniformly spaced
grid of $16\times16$ test points within the cell.  We use Mangle2 to
test if points fall inside the mask \citep{Mangle}.  This provides us
with a weight map $w_i=1/f_i$ where $f_i$ is the fractional sampling
for pixel $i$.  Cells that have less than 50\% inclusion in the survey
are removed from the map. The areas of the four fields W1, W2, W3 and
W4 are 57.7, 18.6, 36.8, 19.6 \sqrdeg.  After putting galaxies in the
{\sc Healpix} cells, the number of pixels in the four maps are: 4787, 1592,
3045 and 1651.

With $n_i$ galaxies counted in cell $i$, the over-density is computed
with $\delta_i=n_iw_i/\bar{n} - 1$.  The mean density in a cell,
$\bar{n}$, is computed from all four fields as $\bar{n}=\sum_i w_i n_i
/ \sum_i w_i$.  The variance of $\delta_i$, assuming Poisson
statistics, is $\sigma_i^2=w_i^2/\bar{n}$.  

The clustering of foreground stars can be a significant source of
systematic error on large angular scales \citep{Ross11}.  For the
CFHTLS, we can estimate the stellar contamination rate independently
in each of the four fields and apply local corrections to the galaxy
density.  We measure the contamination rate directly in the W1 and W4
fields by counting the number of targets spectroscopically classified
as stars in the VIPERS sample.  We then extrapolate these rates to the
W2 and W3 fields by computing the fraction of sources photometrically
classified as stars and then scaling.  

The total counts in a cell broken down into stars and galaxies is
given by $N_{observed}=N_{\star}+N_{galaxy}$, and the stellar
contamination fraction is $f_\star = N_{\star}/N_{observed}$.  The values
we derive are listed in Table \ref{table:stars}.  We apply the
correction in the following way:
$\delta_{i,corr}=\delta_i/(1-f_\star)$ and
$\sigma_{i,corr}^2=\sigma_i^2/(1-f_\star)^2$ \citep{Huterer01}.  The
effect on the amplitude of the power spectrum is $\sim5$\%.

A fraction of galaxies will also be misclassified as stars and removed
from the sample.  However, as long as the sample is representative of
the full population, the power spectrum measurement will not be
biased.  This may not be the case in reality since misclassified
galaxies may preferentially represent a subclass with a different
power spectrum amplitude.  We do not investigate this correction here.

\begin{table}
\caption{Samples \label{table:samples}}
\begin{tabular}{lrrrr}

\hline
Sample & $\zbar$ & $N_{\rm spec}$ & $n_{{\rm phot}}$ & $\bar{n}/$\sqrdeg \\
\hline
SV: VIPERS-like       & 0.70  & 13191 & 1870617 & 14099\\
S6: $0.5<\zphot<0.6$  & 0.56  &  2548 &  340611 &  2567\\
S7: $0.6<\zphot<0.8$  & 0.69  &  4969 &  613643 &  4628\\
S8: $0.8<\zphot<1.0$  & 0.84  &  3030 &  416897 &  3142\\
\hline
\end{tabular}

\caption{Star contamination fractions \label{table:stars}}
\begin{tabular}{crrrr}

\hline
Sample & W1 & W2 & W3 & W4 \\
\hline
SV     & 0.019 & 0.056 & 0.020 & 0.044 \\
S6     & 0.013 & 0.030 & 0.010 & 0.017 \\
S7     & 0.015 & 0.042 & 0.015 & 0.032 \\
S8     & 0.017 & 0.055 & 0.019 & 0.048 \\
\hline
\end{tabular}
\end{table}

\subsection{Redshift distribution}
\label{sec:zdist}
We use the VIPERS spectroscopic redshift catalogue (internal release,
version 1.1) to calibrate the redshift distribution of the photometric
samples, see Table \ref{table:samples}.  In total, we use 13191
galaxies from VIPERS including 6516 from the W1 field and 6675 from
the W4 field.  All targets that meet the photometric selection
criteria with secure redshifts are used.  We select based on the
quality flag \texttt{zflag}. Reliable redshifts have $\texttt{zflag}
~{\rm modulo}~ 10\ge2$, and we take $\texttt{zflag}~{\rm in}~
\{2..9\}$ (galaxy type), $\{12..19\}$ (AGN type) and $\{22..29\}$
(serendipitous detections).  The flag also has a fractional part
indicating agreement with the photometric redshift on a scale from 1
to 5, where 5 indicates good agreement (within $1\sigma$).

We estimate the redshift distribution from the histogram of
spectroscopic redshifts with a bin size of $\Delta {\rm z}=0.05$, see
Fig. \ref{fig:zdist}.  We use the histograms directly in the analysis
with linear interpolation between bin centres.  The redshift
distribution of the W1 and W4 samples are remarkably similar despite
that these fields are well separated on the sky.  We use the
distributions from the two fields to test the impact of cosmic
variance on our results.  As we will conclude in Section
\ref{sec:sims}, small perturbations to the redshift distribution do
not strongly impact the results.

The selection function of the VIPERS survey is not uniform with target
apparent flux.  There are two sampling rates that we consider: first,
the fraction of potential targets that are selected for observation,
and second, the fraction of observed targets that give a successful
redshift measurement.  The first distribution is nearly uniform; the
VIMOS spectrograph can place slits on $\sim$40\% of the potential
targets and this fraction is found to be independent of the magnitude
of the source.  However, we do find that the fraction of targets that
have a measured redshift with qualities meeting our criteria drops
from 100\% at $i=19$ to 50\% at $i=22.5$, the flux limit of the
survey.  This trend may be corrected for by weighting the contribution
of each galaxy in the redshift distribution by the inverse of the
sampling rate.  However, we find that the correction has a negligible
effect on the distribution. Weighting the galaxies shifts the mean
redshifts of the samples by less than $\Delta{\rm z}=0.01$.  We also
confirm that lowering the quality threshold of the spectroscopic
sample to $\texttt{zflag}>=1.5$, which adds 11\% additional sources,
does not significantly alter the distribution.  In Section
\ref{sec:sims} we consider the effect of shifting the mean redshift by
$\Delta{\rm z}=0.05$ to provide an overly-conservative check on the
effect of uncertainties in the redshift distribution.

\section{Angular Power Spectrum}
From galaxy counts in an image we may infer the projected over-density
of galaxies on the sky, $\delta(\n) = \int _0 ^\infty
\delta_{3D}(\n,r) \phi(r) r^2 dr$.  Typically this is an integration
through a broad slice in redshift defined by a photometric galaxy
selection function or simply by the limiting flux of the survey. We
expand the density field in spherical harmonics and express the power
in mode $l$ by the spectrum $\Cl$.

We may write the angular power spectrum as a projection of the
three-dimensional power spectrum, $P_k=\langle\left|\delta_{3D,k}
\right|^2\rangle$.  On large scales, the power spectrum evolves with
the linear growth factor, $D_1(\z)$.  We scale the power spectrum
taken at the median redshift of the sample, $\zbar$, as
$P_k(\z)=\left[D_1(\z)/D_1(\zbar)\right]^2 P_k(\zbar)$.  This gives,
\begin{equation}
\Cl = \frac{2}{\pi} \int  \left[\int \phi(r) D_1(\z)/D_1(\zbar) j_l(kr)  r^2 dr\right]^2 P_k(\zbar)  \frac{dk}{k},
\end{equation}
where $r$ is comoving distance.  In the small angle approximation, the
spherical Bessel function can be approximated as $j_l(x) =
\sqrt{\frac{\pi}{2l+1}}\delta(l+\frac{1}{2}-x)$, expressed with a
Dirac delta function, and we find Limber's equation:
\begin{equation}
\label{eq:limber}
\Cl =  \int g_l(k) P_k\frac{dk}{k}.
\end{equation}
The projection kernel, $g_l(k)$, is given by,
\begin{equation}
\label{eq:kernel}
g_l(k) = \frac{1}{l+1/2}\left[r^2\phi(r) D_1(\z)/D_1(\zbar)\right]^2~~{\rm at}~~r = \frac{l+1/2}{k}.
\end{equation}
A correction may be added to account for redshift space distortions
although it is sizable only on large scales at $l<50$ that we are not
sensitive to here \citep{Thomas11,Ross11a}.  We may now approach the
deprojection problem as a deconvolution of Limber's equation.

\begin{figure}
\includegraphics[scale=1]{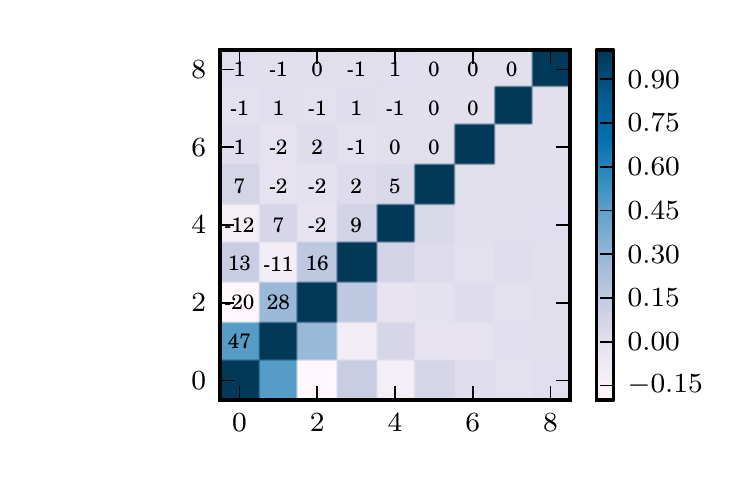}
\caption{ The correlation matrix for $P_k$ estimated from the
  VIPERS-like sample.  Elements of the matrix are labelled with the
  per-cent correlation.  Although the window functions overlap
  significantly, the bins are nearly statistically independent by
  construction.\label{fig:covmat}}
\end{figure}

\section{Power spectrum estimator}
\label{mlhood}
On large scales, the galaxy density may be described by a Gaussian
random field and the distribution is fully characterised by its
variance.  With this assumption, the likelihood function of the
observed overdensities on the sky may be written explicitly.  We order
the $m$ pixels of the density map and form a data vector,
$\x=[\delta(\n_0),\delta(\n_1),...,\delta(\n_{m-1})]$, and write the
covariance of the data as $C_{ij} = \langle x_i x_j \rangle$.  The
likelihood function is,
\begin{equation}
\label{eq:like}
L = \frac{1}{\sqrt{(2\pi)^m \det\C}}\exp\left[-\frac{1}{2}\x^T\Cinv\x\right]
\end{equation}
 The covariance between the overdensity in pixels $i$ and $j$
 separated by an angle $\theta_{ij}$ is given by the sum of the signal
 and the noise components,
\begin{equation}
C_{ij} = \sum_l \frac{2l+1}{4\pi} \Pl(\cos \theta_{ij}) B_l^2 \Cl + N_{ij}
\end{equation}
where $N_{ij}$ is the noise covariance matrix and $\Pl$ are Legendre
polynomials.  The noise matrix is taken to be diagonal with Poisson
elements given by $N_{ii}=w_i^2/\bar{n}$.  The finite resolution of
the pixelised map attenuates the power spectrum by the pixel window
function, $B_l$, which depends on the pixel geometry \citep{Healpix}.

We now derive a power spectrum estimator that maximises the likelihood
function, $L$.  The quadratic form was introduced by \citet{Tegmark97}
and explicit derivations have been given in \citet[Ch. 11]{Dodelson},
\citet{Dahlen08} and \citet{Bond98}.  We give an overview here, since
many variations exist.

We denote the set of parameters to be estimated by the vector $\lam$.
For our study, $\lambda_i$ will represent a bin of the power spectrum.
We begin with an initial estimate, $\lam^{(0)}$, and intend to use an
optimisation algorithm to find a better estimate, $\hat{\lam}$, that
maximises the likelihood function. With the assumption that $\ln L$
has a quadratic form near the peak, we may apply the Newton-Raphson
root-finding method to move toward the peak of the likelihood function
\citep{numericalrecipes}, with,
\begin{equation}
\label{eq:newton}
\hat{\lam} = \lam^{(0)} - \frac{\dd \ln L / \dd \lam}{\dd^2 \ln L / \dd \lam^2}\bigg|_{\lam^{(0)}}.
\end{equation}
This expression may be used iteratively to locate the peak.  

We evaluate the derivative terms in Appendix \ref{appendix}, and
now simply state the final result for one iteration step:
\begin{equation}
\hat{\lambda}_i = \frac{1}{2} \sum_{j}A_{ij}  \left\{  \x^T {\bf E}_j \x  - \Tr \left( {\bf E}_j {\bf N}\right)\right\}
\label{eq:est}
\end{equation}
\begin{equation}
\label{eq:deriv}
{\bf E}_j = {\bf C}^{-1} \dCdj {\bf C}^{-1}
\end{equation}
The matrix ${\bf A}$ is a mixing matrix that sets the normalisation
and may be specified to form linear combinations of the bin estimates.
We will use this matrix to shape the window functions.  The second
term in Eq. \ref{eq:est} subtracts the noise bias.

We see that the estimator weights the data by their covariance:
$\Cinv{\bf x}$.  This approach has the favourable property that spatial
modes contribute to the measurement with an inverse-variance weight.
The weighting also appropriately `tapers' the map near the mask
boundary giving compact window functions in harmonic space.

We have not yet specified the parameter vector $\lam$.  We set $\lam$
to bins of the three-dimensional power spectrum and evaluate the
derivative matrix in Eq. \ref{eq:deriv}, as,
\begin{equation}
\frac{\dd C_{ij}}{\dd \lambda_k} \equiv \frac{\dd C_{ij}}{\dd P_k}= \sum_{l=2}^{l_{max}} \frac{2l+1}{4\pi}\Pl(\cos\theta_{ij})B_l^2 g_l(k) \Delta_{\ln{k}}.
\end{equation}
Here, we have replaced the integral in Limber's equation
(Eq. \ref{eq:limber}) with a discrete sum over $\ln{k}$ with
logarithmic bin width $\Delta_{\ln{k}}$.

The expectation of the estimate is given by,
\begin{equation}
\langle \hat{\bf \lambda} \rangle = {\bf A}{\bf F} {\bf \lambda},
\end{equation}
where we have introduced the Fisher matrix,
\begin{equation}
F_{ii'}=\frac{1}{2} \Tr \left( {\bf C}^{-1} \frac{\dd{\bf
      C}}{\dlambda_i} {\bf C}^{-1} \frac{\dC}{\dlambda_{i'}}
  \right).
\label{eq:fish}
\end{equation}
The variance of the estimate is,
\begin{equation}
\label{eq:var}
{\rm Var}(\hat{\bf \lambda},\hat{\bf \lambda}) = {\bf A} {\bf F} {\bf A}^T.
\end{equation}
and the window functions are ${\bf W} = {\bf A}{\bf F}$.

The inverse of the Fisher matrix represents the minimum variance that
we may hope to achieve on $\hat{\bf \lambda}$.  With ${\bf A}={\bf
  F}^{-1}$ we see that we have an estimator that is optimal in the
sense that it is unbiased and has minimum variance \citep{Tegmark97}.
This approach may not be practical however, because the Fisher matrix
is often singular or numerically ill-conditioned.  Intuitively, this
reflects the fundamental limit that we cannot probe the power spectrum
at scales smaller than $\Delta\ell\sim(\Delta\theta)^{-1}$ where
$\Delta \theta$ is the angular size of the survey.  

Instead, we choose ${\bf A}$ with the aim of diagonalising the
covariance matrix. By factoring the Fisher matrix as ${\bf F}={\bf
  M}{\bf M}^T$, we can set ${\bf A} = {\bf M}^{-1}$.  The covariance
matrix is now ${\rm Var}(\hat{\bf \lambda},\hat{\bf \lambda}) = {\bf
  M}^{-1} {\bf F} {{\bf M}^{-1}}^T$.  In practice, we compute ${\bf
  M}$ as the square root of the Fisher matrix using a singular value
decomposition (SVD) method.  We also rescale ${\bf M}$ to normalise
the window functions such that $\sum_jW_{ij}={\bf 1}$.  This approach
was shown by \citet{Tegmark97} to result in sharper window functions
than the common choice for ${\bf A}$, a diagonal matrix with
$A_{ii}=\left[ \sum_j F_{ij} \right]^{-1}$.

We find that the matrix ${\bf M}$ is ill-conditioned when the window
functions are broad, especially for the SV sample which has a wide
redshift distribution.  To find a stable inversion, we use a
pseudo-inverse technique by keeping only the largest singular
values. The consequence of using a pseudo inverse is that the
covariance matrix will not be perfectly diagonalised.  The covariance
matrix for the VIPERS estimate after carrying out this operation is
shown in Fig. \ref{fig:covmat}.  The choice of how many modes to keep
in the pseudo-inverse affects the shape of the window functions and
the scales that are probed. We find that the smaller singular values
probe large scales, in the same fashion as in the SVD analyses by
\citet{Eisenstein01} and \citet{Maller05}.  We set the scale ensuring
that the resulting window functions are positive and reach the largest
scales available to the survey.

The estimate and covariance model depend on the chosen fiducial
cosmology through the form of the likelihood function and the
projection kernel.  Both the shape and normalisation of the power
spectrum can be important.  Although the normalisation cancels in the
estimator (neglecting the noise term), it is important for the Fisher
matrix and covariance.  For these reasons, maximum likelihood
estimators are often applied iteratively to arrive at consistent
results.  We explore these dependencies with simulations in the next
section.

\begin{figure}
\includegraphics[scale=1]{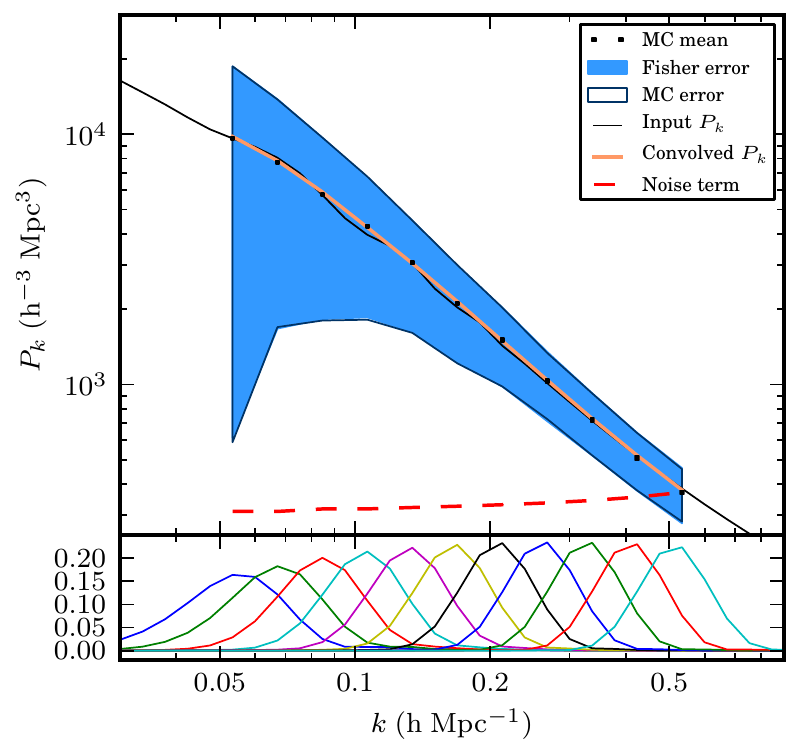}
\caption{ The top frame shows the recovered power spectrum from the
  mean of 1000 independent Gaussian simulations.  The theory is
  convolved with the window functions (plotted at bottom) and we find
  that it matches the measurement to within a few percent.  The error
  bars computed analytically from the Fisher matrix (shaded area)
  agree with the distribution of MC runs (outline).  The noise term in
  Eq. \ref{eq:est} is shown as a dashed curve. \label{fig:sim}}
\end{figure}

\begin{figure}
a)\includegraphics[scale=1]{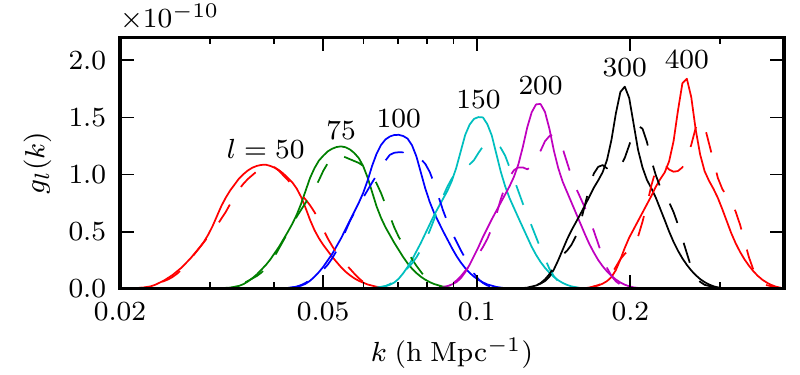}
b)\includegraphics[scale=1]{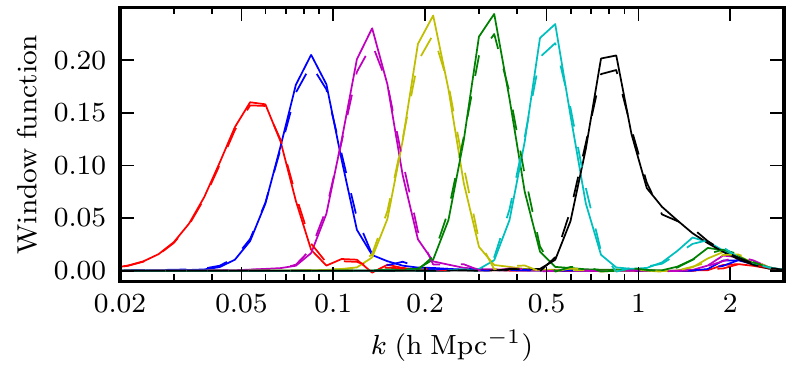}
\caption{In the top panel we plot the projection kernels, $g_l(k)$.
  We compare the kernels derived from the W1 redshift distribution
  only (solid curves) and those from W4 only (dashed curves).  Lower
  panel: the window functions found for the 21\sqrdeg~simulation field
  are plotted. To check robustness we again compare the results
  derived from the W1 and W4 fields separately.  The second peak in
  the window functions at $k>1\hmpc$ arises from the pixel scale of
  the map; beyond $k=2\hmpc$ the window functions rapidly drop to
  0. \label{fig:win}}
\end{figure}

\begin{figure}
\includegraphics[scale=1]{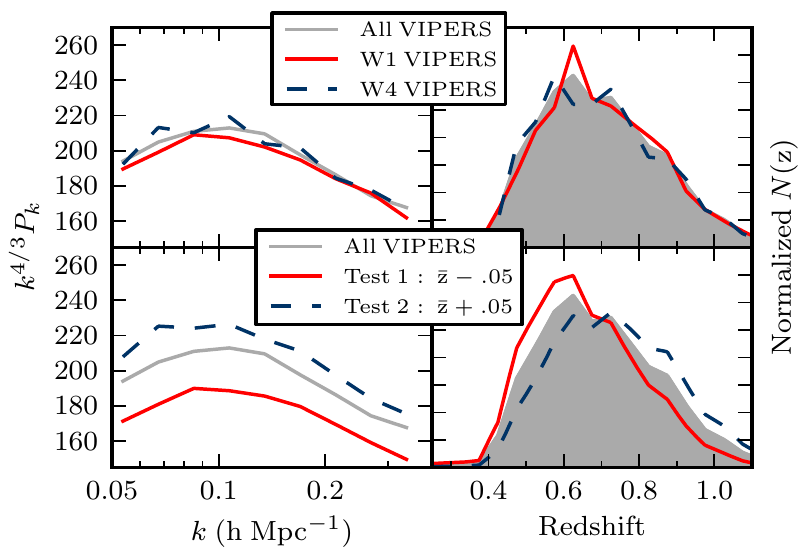}
\caption{The influence of the assumed redshift distribution on the
  deprojection.  Top panels: the simulations are analysed with the W1
  and W4 redshift distributions.  Bottom panels: for an
  overly-conservative test, we modify the VIPERS redshift distribution
  to adjust the mean redshift by $\Delta\z=\pm0.05$ (labelled Test 1
  and Test 2) leading to systematic shifts in the amplitude of the
  estimated power spectrum of $+6$\% and $-10$\%.  The redshift
  distributions are plotted in the right panels and the derived power
  spectra are on the left. In the top and bottom, the filled grey
  distribution represents the redshift distribution of the full VIPERS
  sample.\label{fig:zdisttest}}
\end{figure}

\begin{figure}
a)\includegraphics[scale=1]{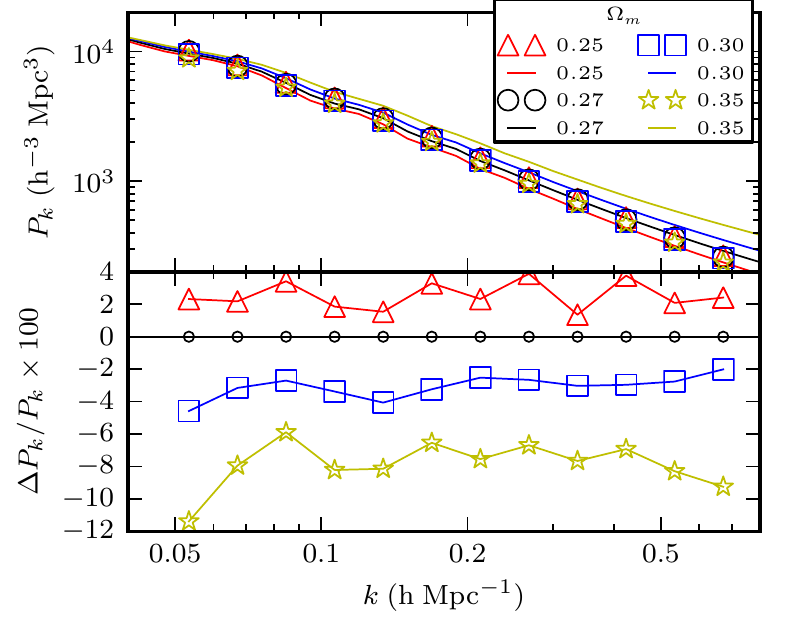}
b)\includegraphics[scale=1]{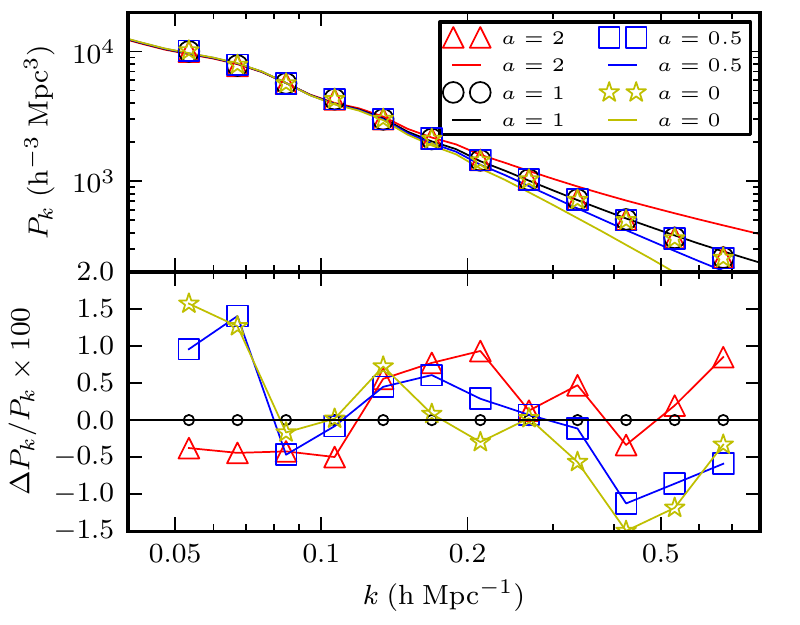}

\caption{The sensitivity of the estimator to the assumed
    fiducial model.  (a) We vary $\Omega_m$ keeping other parameters
    fixed.  The top frame shows the fiducial power spectra (solid
    lines) and the points mark the derived power spectrum
    measurements.  The bottom frame shows the percent differences from
    the reference model for each trial.  The correct shape is
    recovered, but there is a shift in the amplitude of the estimate
    due to the geometric dependence of the projection kernel on the
    cosmology.  (b) We modulate the small-scale amplitude of the
    fiducial power spectrum with an interpolation parameter $a_{nl}$;
    $a_{nl}=0$ and 1 correspond to the linear and Halofit models,
    respectively.  We find that the derived power spectrum is not
    sensitive to the small-scale amplitude.
 \label{fig:nl}}
\end{figure}

\section{Simulations}
\subsection{Gaussian realisations}
\label{sec:sims}
As a test of the method, we estimate the power spectrum for Gaussian
realisations of the projected density field.  The simulations are
constructed using a CAMB power spectrum with the Halofit model
\citep{camb,Smith03} and projected through the VIPERS redshift
distribution.  With these $\Cl$, we use {\sc Healpix} {\sc Synfast} to generate
simulated density maps.  We produced 1000 independent maps with a
resolution of 7\arcmin~($\nside=512$).  We add noise fluctuations by
drawing from a Gaussian distribution assuming a variance given by
Poisson statistics with $\bar{n}=50$ galaxies/cell. This is a higher
level of noise than we find in the CFHTLS catalogue.  For the geometry
of the mock survey, we use the actual survey mask of the W2 field.  It
includes 1592 pixels covering 21 \sqrdeg.

We compute the Fisher matrix in logarithmic bins from $k=0.01-100$ with
$\Delta_{\log k}=0.05$.  The results are plotted in
Fig. \ref{fig:sim}.  The data points are re-binned to $\Delta_{\log
  k}=0.1$ and plotted from $k=0.06-0.7 \hmpc$.  The corresponding
window functions are shown in the bottom panel.  The data are plotted
at the peaks of the window functions. On large and small scales, the
window functions begin to overlap and converge as the limits set by
the survey geometry are reached.  On small scales, we see a secondary
peak in the window function at $k\sim2 \hmpc$ which arises from the
pixel scale of the map, see Fig. \ref{fig:win}.

We convolve the theory power spectrum with the window functions and
find that the mean of the Monte Carlo runs agrees well within a few
percent.  We expect that the precision is limited by the finite
binning of the Fisher matrix and truncation of the window functions,
but these effects are well below the statistical uncertainties.  The
errors computed analytically from the Fisher matrix agree with the
distribution of Monte Carlo runs to within a few percent.  These
errors are for a single field, and so, we can expect to achieve a factor
of two better with the combination of four fields.

\subsection{Dependence on redshift distribution}
We checked the robustness of the measurement to uncertainties in the
redshift distribution by repeating the analysis with different assumed
distributions.  The simulations were generated with the measured
redshift distribution from the complete VIPERS sample, and we first
re-analysed them with distributions derived from two subsamples, the
redshift distribution of W1 and W4.  The numbers of spectra taken in
the two fields are similar and the spectroscopy covers similar areas,
but, the fields are widely separated on the sky and so any differences
could be attributed to cosmic variance.  We compute the Fisher matrix
using the two distributions and find that the projection kernels and
window functions agree (Fig. \ref{fig:win}).  The bias introduced by a
mismatched redshift distribution is at the percent level, below the
statistical errors.

Additionally, the measured redshift distribution could be inaccurate
due to sampling biases in the VIPERS survey. In Section
\ref{sec:zdist}, we concluded that the uncertainty in the mean
redshift of the distribution, $\zbar$, is known to better than $\Delta
\z=0.01$.  As an overly conservative check, we examine the
consequences of shifting the redshift distribution by $\Delta
\z=\pm0.05$.  This was done by modulating the measured distribution of
the full VIPERS sample by the linear function
$f(\z)=1\pm1.5(\z-\zbar)$.  The modified distributions have
$\zbar_1=0.656$ and $\zbar_2=0.752$, while the original sample has
$\zbar=0.703$, see the lower panel of Fig. \ref{fig:zdisttest}.  We
find that reducing $\zbar$ by 7\% lowered the derived power spectrum
by 10\%.  Increasing $\zbar$ by 7\% increased the power spectrum by
6\%.  This large shift in $\zbar$ would thus lead to a systematic
error in the estimated bias factor at the level of $3-5$\%.

\subsection{Dependence on fiducial cosmology}
The dependence on the fiducial cosmology enters the analysis in two
ways.  First, we rely on the cosmology to model the likelihood
function.  In the maximum likelihood estimator, the data covariance
matrix plays the role of a weight. Modifying the fiducial power
spectrum changes the weighting function and could bias the estimator.
We can expect that assuming the wrong matter density for example,
could bias the estimator and make the variance properties sub-optimal.

The second dependence on the fiducial cosmology is through the
projection kernel.  In the previous section, we discussed how shifting
the redshift distribution affects the amplitude of the power spectrum
estimate.  We can expect that varying the cosmology and the
redshift-distance relation will have a similar effect.

Our Gaussian simulations were constructed using the reference \LCDM~
power spectrum with the Halofit model.  To test the dependence on the
cosmology, we first re-analyse the maps using fiducial power spectra
with different assumed values of the matter density, taking $\Omega_m
= 0.25$, 0.30 and 0.35.  All other parameters were held fixed at the
reference values.  We find that despite assuming the wrong matter
density, we recover the correct shape of the power spectrum from the
mean of 1000 simulation runs to within 2\%, see Fig. \ref{fig:nl},
Panel a.  However, it is clear that the amplitude is strongly biased.
This is due to the dependence of the projection kernels on $\Omega_m$.
This geometric dependence on the background cosmology dominates over
any bias in the estimator due to sub-optimal weighting.  These
findings support an iterative approach.

Next, we check the influence of variations in the amplitude of the
power spectrum at small scales.  The shape of the power spectrum on
small scales has developed with the aid of n-body simulations but it
remains a source of systematic uncertainty. We vary the small-scale
amplitude using an interpolation parameter $a_{nl}$:
\begin{equation}
\tilde{P}_k = P_{k,lin} + \left( P_{k,nl} - P_{k,lin} \right)a_{nl}.
\end{equation}
We test a range of amplitudes with $a_{nl}=\{0,0.5,1,2\}$ ($a_{nl}=1$
gives the Halofit model).  We find that the estimator is remarkably
robust, see Fig. \ref{fig:nl}, Panel b.  The discrepancy introduced by
the variation in the small-scale amplitude is less than $2\%$ on large
scales and it is dominated by numerical uncertainties up to
$k\sim0.2\hmpc$.  This supports the conclusion that using sub-optimal
weights does not significantly bias the result.

\begin{figure*}
\includegraphics[scale=1]{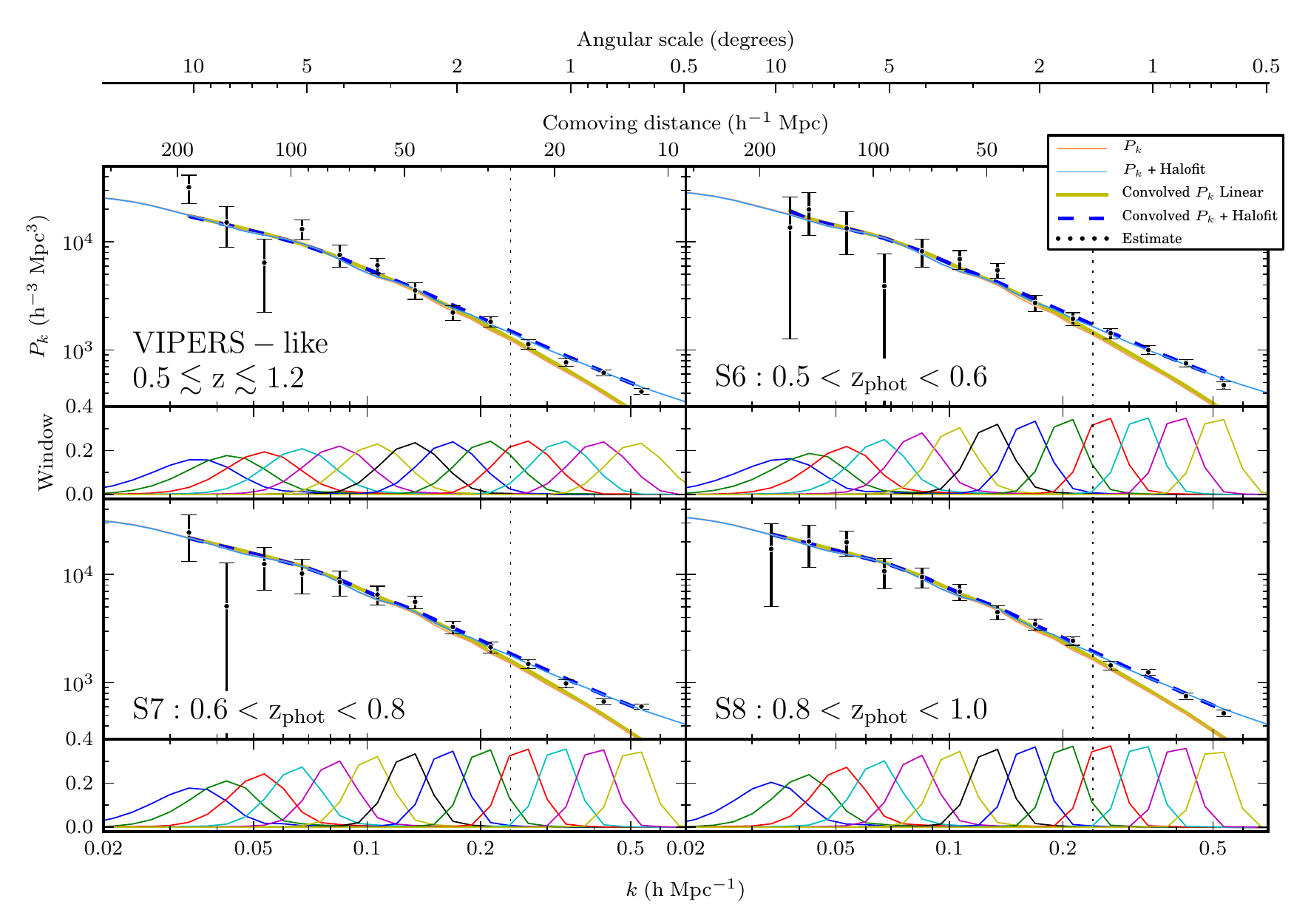}
\caption{The deprojected three-dimensional power spectrum in
  logarithmic bins for the VIPERS-like sample and three photometric
  redshift selected samples: $\rm{z_{phot}}=0.5-0.6$, $0.6-0.8$ and
  $0.8-1.0$. The window function of each band is shown in the panel
  below each plot.  The theoretical power spectrum (both linear and
  Halofit models) is convolved with the window function and
  overplotted with the best-fit linear bias listed in Table
  \ref{table:bias}.  We use the first 9 data points, up to
  $k\simeq0.2\hmpc$ indicated by the vertical dotted line, to estimate
  the linear bias.  The corresponding comoving distance
    and angular scale (at $\z=0.7$) are included as a guide.
\label{fig:results}}
\end{figure*}

\section{Results}
We now carry out the estimation of the power spectrum on the CFHTLS
data for the VIPERS-like sample and three photometric redshift
subsamples, see Fig. \ref{fig:results}.  For the analysis, a fiducial
CAMB power spectrum with the Halofit model is assumed.  The Fisher
matrix is computed with 60 bins logarithmically spaced from $k=0.01 -
10$ with $\Delta\log k=0.05$.  We use a wide $k$-range to map out the
window functions but all these data are not useful for analysis.  We
restrict the study to 13 points from $k=0.03-0.6\hmpc$.  We can go to
smaller scales, although the Gaussian error estimate will not be
appropriate.  We use a bin size of $\Delta\log k=0.1$ which is
appropriate choice considering the width of the survey window
functions.  The plotted error bars are derived from the diagonal
elements of the covariance matrix found computed from
Eq. \ref{eq:var}.

For each field, we compute the normalised quantity from
Eq. \ref{eq:est}, $y_j^k = \frac{1}{2} \left\{ \x^T {\bf E}_j \x - \Tr
\left( {\bf E}_j {\bf N}\right)\right\}$, where $k$ indexes the fields
1-4, along with the Fisher matrix (Eq. \ref{eq:fish}).  These results
are then summed together, and the final combined estimate is computed
by $\hat{\lambda}_i = \frac{1}{2} \sum_{k=1}^4 \sum_{j} A_{ij} y^k_j$
where ${\bf A} = \left(\sum_{k=1}^4 {\bf F}^k\right)^{-1/2}$
normalised such that $\sum_i\left({\bf A}{\bf F}\right)_{ij}={\bf
  1}$. This combination properly weights the data. The covariance of
the estimate for the VIPERS-like sample is shown in
Fig. \ref{fig:covmat}.  At low $k$, neighbouring bins are nearly 50\%
correlated, but the matrix becomes more diagonal at larger $k$.  The
limitation in diagonalising the covariance matrix comes in the
inversion of ${\bf M}={\bf F}^{1/2}$.  This is computed with a
pseudo-inverse method.  The inversion becomes easier for the narrower
photometric redshift slices where a nearly perfect inversion is
possible.  The window functions are sharper for these redshift slices
as well.

We do not run the maximum likelihood algorithm in an iterative
fashion.  The data do not support strong constraints on
\LCDM~parameters alone and we find that beginning with a fiducial
\LCDM~power spectrum, we have a very good fit to the data.  This
indicates that our starting point is already near to the peak of
the (very broad) likelihood function.  However, we do effectively
carry out one iteration of the estimator to find the galaxy bias and
set the amplitude of the fiducial power spectrum. This is necessary
because the estimator and covariance do not simply scale linearly with
amplitude in the presence of noise.  A second run allows us to set the
amplitude of the fiducial power spectrum ensuring that the error
estimate is correct.

We compute a one-parameter fit to estimate the galaxy bias on linear
scales.  We restrict this fit to the first 9 points at $k<0.2\hmpc$.
Given the $P_k$ measurements in vector ${\bf d}$ and the convolved
\LCDM~model in vector ${\bf m}$, we find the amplitude, $a$, that
maximises the likelihood function, $\ln L = -1/2\left({\bf d}-a{\bf
  m}\right)^T\Cinv_{kk'}\left({\bf d}-a{\bf m}\right)$.  The solution is
given by,
\begin{equation}
a=\frac{{\bf d}^T\Cinv_{kk'}{\bf m}}{{\bf m}^T\Cinv_{kk'}{\bf m}}
\end{equation}
with variance $\sigma^2_a=\left({\bf m}^T\Cinv_{kk'}{\bf
  m}\right)^{-1}$.  The resulting values of the galaxy bias are listed
in Table \ref{table:bias}, where we have assumed a value of
$\sigma_8=0.8$.  The bias increases with redshift as expected for a
flux limited survey.  In fact, the amplitude of the power spectrum is
not seen to change with increasing redshift, indicating that the
evolution of the growth factor and the galaxy bias factors
approximately cancel.  We find an approximately constant error on the
bias factor in each redshift range; this is simply due to the fact
that the amplitude of the fiducial power spectrum (from which the
errors are derived) is approximately constant.

Also in Table \ref{table:bias}, we list the $\chi^2$ values at the
best fit.  The number of degrees of freedom is approximately 8.  The
$\chi^2$ values are lower than expected, specifically for S8 for which
we find $\chi^2=2$.  Formally, the probability of finding $\chi^2\le2$
with 8 degrees of freedom is $0.019$.  This could indicate that the
covariances and, consequently the error bars, are over-estimated
for this sample.

\begin{figure}
\includegraphics[scale=1]{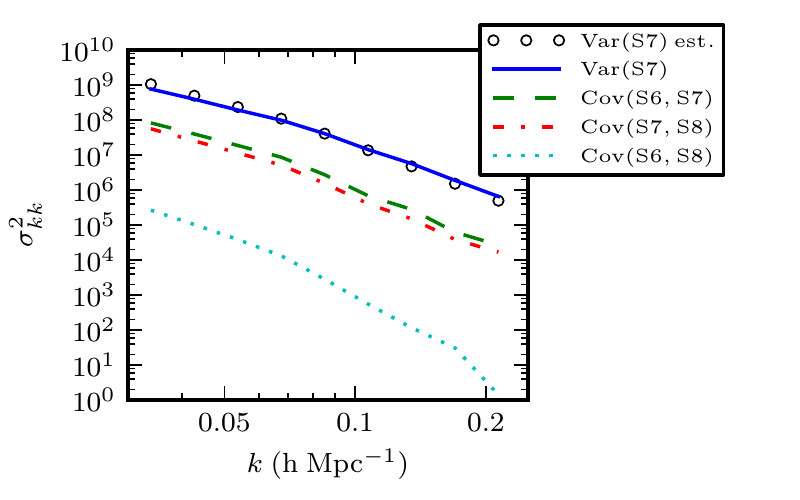}
\nolinebreak\hspace{-27mm}\includegraphics[scale=0.7]{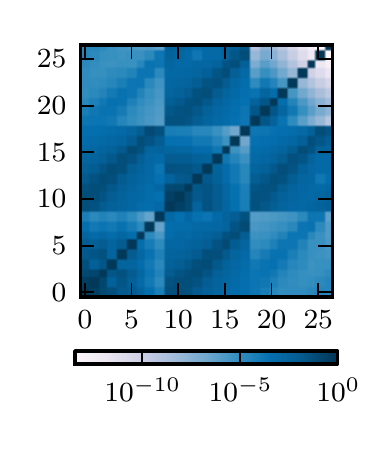}
\caption{The covariance between the photometric redshift samples is
  plotted.  The circle markers show the full computation of the
  variance of the S7 slice that accounts for the
  survey geometry.  The solid curves show the analytic Fisher
  approximation in the full-sky limit (Eq. \ref{eq:cov}).  Inset is
  the full correlation matrix for the three samples.
 \label{fig:samplecov}}
\end{figure}

\begin{table}
\caption{Best-fit galaxy bias \label{table:bias}}
\begin{tabular}{ccr}

\hline
Sample & $b_g$ & $\chi^2$ \\
\hline
VP: VIPERS-like       & $1.38 \pm 0.05$ & 7.2 \\ 
S6: $0.5<\zphot<0.6$  & $1.36 \pm 0.05$ & 7.1 \\
S7: $0.6<\zphot<0.8$  & $1.52 \pm 0.05$ & 5.7 \\
S8: $0.8<\zphot<1.0$  & $1.68 \pm 0.05$ & 2.0 \\

\hline
\end{tabular}

\caption{Marginalised parameter estimates \label{table:param}}
\begin{tabular}{ccr}
\hline
$\Omega_m$ & $0.30 \pm 0.06$ \\
$b_{S6}$ & $1.39 \pm 0.08$ \\
$b_{S7}$ & $1.55 \pm 0.08$ \\
$b_{S8}$ & $1.72 \pm 0.10$ \\

\hline
\end{tabular}
\end{table}

\section{Parameter estimation}
Before we may carry out a joint analysis, we must estimate the
covariance between the overlapping photometric samples.  We will
estimate the covariance of the estimators with a Fisher matrix
approach in the full-sky limit and then rescale to find the errors
for our survey geometry.

We compute the covariance between two samples labelled $A$ and $B$.
To simplify the expression, we write the product of the kernel with
the beam and integration step as,
$\tilde{g}_l(k)=g_l(k)B_l^2\Delta_{\ln k}$ and the sum of the signal
and noise covariance as, $\tilde{\Cl}=\Cl
B_l^2+\frac{\Delta\Omega}{\bar{n}}$.  It is more convenient to use the
harmonic space representation, and we switch the data vector from
$x_i$ to $a_{lm}$.  The covariance matrix is diagonal:
$C_{lm;l'm'}=\delta_{ll'}\delta_{mm'}\tilde{\Cl}$.

 A component of the Fisher matrix
for a single sample $A$ for two power spectrum bins, $k$ and $k'$ is,
\begin{equation}
F_{A,kk'}=\sum_l\frac{2l+1}{2} \tilde{g}^{A}_l(k)\tilde{g}^{A}_l(k')\left(\tilde{\Cl}^A\right)^{-2}
\end{equation}
We can write the quadratic estimator for the power spectrum as,
\begin{equation}
\hat{P}^A_k = \frac{1}{2}\sum_{k'}F_{A,kk'}^{-1} \sum_{lm}  a_{lm}^2\tilde{g}^A_l(k')\left(\tilde{\Cl}^A\right)^{-2}
\end{equation}

We find the covariance between the two sample estimates to be,
\begin{eqnarray}
{\rm Cov}(\hat{P}_k^A,\hat{P}_{k'}^B)&=&  \frac{1}{f_{sky}} \sum_{h,i,j} F_{A,ki}^{-1}F_{B,ij}^{-1}\times \nonumber \\
&&\sum_l\frac{2l+1}{2} \tilde{g}^{A}_l(j)\tilde{g}^{B}_l(k') \left( \frac{\tilde{\Cl}^{AB}} {\tilde{\Cl}^A \tilde{\Cl}^B} \right)^2\label{eq:cov}
\end{eqnarray}
We scale by the fractional sky coverage of the survey, $f_{sky}$,
which approximately accounts for the number of modes that may be
probed.  The small survey size also broadens the window functions,
which we account for in the covariance with: $\C_{kk'}' = {\bf W^A}
\C_{kk'} {\bf W^B}^T$.

The variances for one sample computed with Eq. \ref{eq:cov} in the
full-sky limit match well with the full computation of the Fisher
matrix (Eq. \ref{eq:var}), see Fig. \ref{fig:samplecov}. Although, we
find that the full-sky computation underestimates the variance by a
factor of $\sim 2$.  This is not surprising since we have neglected
the precise survey geometry.  As a correction, we rescale the estimate
to match the variance in the S7 slice.  In Fig. \ref{fig:samplecov} we
also show the analytic estimates of the covariances between the three
redshift slices that we may now use to perform a joint likelihood
analysis.

\begin{figure}
\includegraphics[scale=1]{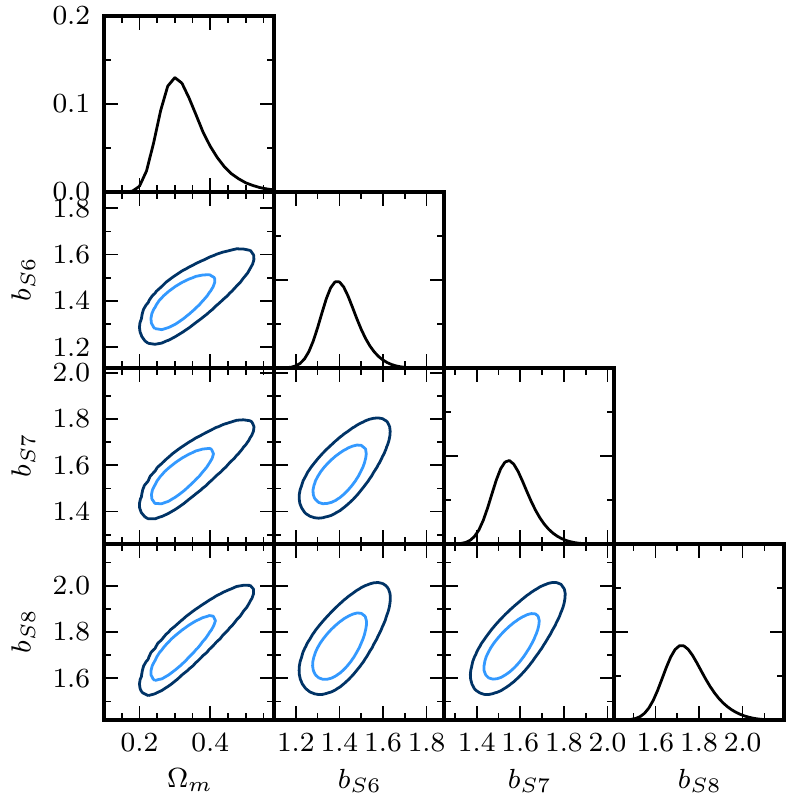}
\caption{The joint likelihood surfaces of $\Omega_m$ and the bias
  parameters for the three photo-z samples ($b_{S6}$, $b_{S7}$, $b_{S8}$). The
  inner and outer contours indicate the 68\% and 95\% confidence
  level. The marginalised likelihoods of each parameter are listed in
  Table \ref{table:param}.  The fit is limited to the first 9 data
  points, giving $k_{max}=0.2\hmpc$.\label{fig:like}}
\end{figure}

Using the sample covariances, we jointly estimate the linear galaxy
bias factors of the three photometric redshift slices, S6, S7 and S8,
labelled as $b_{S6}$, $b_{S7}$ and $b_{S8}$ along with $\Omega_m$.
All other \LCDM~parameters are held fixed and we set $\sigma_8=0.8$.
We compute the likelihood of a model with the full covariance matrix.
The fit is limited to the first 9 datapoints of each sample, giving a
maximum $k$ of $k_{max}=0.2\hmpc$.  We exhaustively evaluate the
likelihood over the 4-dimensional parameter grid.  Views of the
likelihood surface, marginalised over pairs of parameters are shown in
Fig. \ref{fig:like}.  The marginalised constraints are listed in Table
\ref{table:param} with 68\% confidence intervals.

The joint analysis prefers a slightly higher value of $\Omega_m$,
$0.30\pm0.06$, versus the fiducial model with 0.272.  Due to the
correlations between parameters, this results in higher values of the
galaxy bias factors than were found with the fiducial model fixed
(Table \ref{table:bias}).

\section{Conclusions}
The CFHTLS-Wide fields probe a significant cosmological volume at
redshifts not reached by other galaxy surveys to date.  We use the
projected density field from photometric redshift samples to constrain
the real-space power spectrum and derive constraints on the matter
density and linear galaxy bias factors.  These results are made
possible by precise knowledge of the redshift distributions provided
by preliminary results from the VIPERS survey.

The primary advantage of computing the power spectrum directly from
the angular distribution, instead of using conventional spherical
harmonics $\Cl$ is that we may construct window functions in Fourier
space.  By optimising this, we achieve sharper constraints on the
power spectrum than when we are limited to $\ell$ bands.  This
approach comes with the cost that we must adopt a fiducial power
spectrum.  We showed that using the wrong fiducial power spectrum,
although leading to sub-optimal weights, does not significantly bias
the estimate.  This is true even on small scales, and we can
effectively use this method to deconvolve small and large scales in
Limber's equation. Residual systematic error on the derived power
spectrum is at the 1\% level, well below the sensitivity of the
measurement.

The deprojection does strongly depend on the assumed redshift
distribution of the galaxy sample as well as the cosmology used to
compute the redshift-distance relation.  The cosmology dependence of
the measurement makes the interpretation difficult, but to a first
approximation, only the amplitude is affected; the shape of the power
spectrum is recovered correctly.  Thus, a converging iterative
procedure can be implemented by updating the fiducial model and
repeating the analysis.

There is a degeneracy between a shift in the assumed redshift
distribution and the cosmological model.  This is unavoidable when
studying a field in projection.  However, the constraints on the
redshift distribution can always be improved with further observation.
In our analysis, from the sampling biases present in the VIPERS
spectroscopy, we estimate the uncertainty in the mean redshift to be
at the 1\% level.  Thus, we do not expect a strong systematic error in
the derived galaxy bias parameters.  We do note that the observed
trend of low $\chi^2$ values for the best-fit models in the higher
redshift samples can arise if the covariance is overestimated.  This
could be a weak hint that the true mean redshift is lower than what we
assume or that a modification is needed in the fiducial cosmology.

Recently, the galaxy bias was measured from the CFHTLS-Wide fields in
the context of the halo model by \citet{Coupon11}.  Our final two
photometric redshift bins, S7 and S8, correspond with samples
constructed by Coupon \etal so we are able to compare the resulting bias
values.  Coupon \etal constructed volume-limited samples using luminosity
cuts resulting in a selection of brighter galaxies, thus we may expect
their bias values to be larger.  The halo model constraints of
\citet{Coupon11} give for S7, $b_g=1.44\pm0.01$, and for S8,
$b_g=1.79\pm0.03$.  These values have been scaled by 1.03 to transform
from a cosmology with $\Omega_m=0.25$ to $\Omega_m=0.272$ which is
assumed here.  Our value of $b_g$ for the S7 sample is higher, while
for the S8 sample it is lower, although both are in agreement with
Coupon \etal within the 2$\sigma$ confidence limit.  The measurements are
based on different physical scales (Coupon \etal restrict the correlation
function to angular scales $<1.5\degr$) and different model
assumptions have been used.  Thus it is reasonable to consider the
measurements as independent estimates.

Our results provide a preliminary look at the large-scale structure
field probed by the VIPERS colour selection and demonstrate the
strengths of the VIPERS sample for clustering studies at $\z>0.5$.  We
anticipate promising results with the full VIPERS spectroscopic
sample.

\section*{Acknowledgements}
We are grateful to the VIPERS team for supporting this project.  In
particular, we thank Lauro Moscardini for suggestions that improved
the work.  We thank Jian-Hua He for carefully reading the manuscript.
The computational methods used were inspired by Istv\'{a}n Szapudi's
{\sc mlhood} code.  Our results are derived with {\sc CosmoPy}
(\href{http://www.ifa.hawaii.edu/cosmopy}{www.ifa.hawaii.edu/cosmopy})
and {\sc Healpix} with {\sc Healpy}
(\href{http://healpix.jpl.nasa.gov}{healpix.jpl.nasa.gov},
\href{http://code.google.com/p/healpy}{code.google.com/p/healpy}).  We
acknowledge the support of INAF through a PRIN 2008 grant.  AP and KM
have beed supported by the research grant of the Polish Ministry of
Science Nr N N203 51 29 38. A part of this work was carried out within
the framework of the European Associated Laboratory ``Astrophysics
Poland-France.''  Based on data obtained with the European Southern
Observatory Very Large Telescope, Paranal, Chile, program 182.A-0886.
Based on observations obtained with MegaPrime/MegaCam, a joint project
of CFHT and CEA/DAPNIA, at the Canada-France-Hawaii Telescope (CFHT)
which is operated by the National Research Council (NRC) of Canada,
the Institut National des Sciences de l'Univers of the Centre National
de la Recherche Scientifique (CNRS) of France, and the University of
Hawaii. This work is based in part on data products produced at
TERAPIX and the Canadian Astronomy Data Centre as part of the
Canada-France-Hawaii Telescope Legacy Survey, a collaborative project
of NRC and CNRS.

\appendix

\onecolumn


\section{Quadratic estimator}
\label{appendix}
In Section \ref{mlhood}, we reasoned that we would apply the
Newton-Raphson root-finding algorithm (Eq. \ref{eq:newton}) to locate
the peak of the likelihood function (Eq. \ref{eq:like}).  We now
continue and evaluate the derivatives of the likelihood function,
$\frac{\dd \ln L}{\dlambda_i}$ and $\frac{\dd^2 \ln
L}{\dlambda_i\dlambda_j}$.  We assume that the covariance of the data
depends linearly on the parameters: $C_{ij}=\sum_k
P_{k,ij}\lambda_{k}+N_{ij}$.


The first derivative term is,
\begin{eqnarray}
\frac{\dd \ln L}{\dlambda_i} &=& \frac{\dd \ln \det\C}{\dlambda_i} + \x^T\frac{\dd \Cinv}{\dd \lambda_i}\x\\
&=&{\rm Tr}\left(\Cinv\dCdi\right) - \x^T\Cinv\dCdi\Cinv\x
\end{eqnarray}
Two identities have been used: $ \ln\left( \det \C\right)=\Tr\left(\ln \C\right)$ and $\frac{\dd\C^{-1}}{\dlambda}=-\Cinv\frac{\dd\C}{\dlambda}\Cinv$.  The second derivative, or curvature, is,
\begin{equation}
\frac{\dd^2 \ln L}{\dlambda_i\dlambda_j} = -\Tr\left(\Cinv\dCdi\Cinv\dCdj\right)+\nonumber 2\x^T\Cinv\dCdi\Cinv\dCdj\Cinv\x
\end{equation}
We neglect the second derivative terms.  To simplify, we replace the
curvature by its average over an ensemble of realizations of the data.
This is known as the Fisher matrix,
\begin{equation}
F_{ij}\equiv\frac{1}{2}\langle\frac{\dd^2 \ln L}{\dlambda_i\dlambda_j}\rangle=\frac{1}{2}\Tr\left(\Cinv\dCdi\Cinv\dCdj\right)
\end{equation}

Inserting these expressions into Eq. \ref{eq:newton}, we find that one
iteration step in the Newton-Raphson algorithm is given by,
\begin{eqnarray}
\hat{\lambda}_i&=&\lambda^{(0)}_i - \sum_j \left(\frac{\dd^2 \ln L}{\dlambda_i\dlambda_j}\right)^{-1}\frac{\dd \ln L}{\dlambda_j}\\
&=&\lambda^{(0)}_i + \sum_j \left(\frac{\dd^2 \ln L}{\dlambda_i\dlambda_j}\right)^{-1}\left(\x^T\C^{-1}\dCdi\C^{-1}\x^T-\Tr\left(\Cinv\dCdi\right)\right)\\
&=&\lambda^{(0)}_i + \frac{1}{2}\sum_j F_{ij}^{-1} \left(\x^T\C^{-1}\dCdi\C^{-1}\x-\Tr\left(\Cinv\dCdi\right)\right)\label{eq:a}
\end{eqnarray}
The terms on the right are computed with the parameter set
$\lambda^{(0)}$.  We may simplify further by rewriting the trace term with,
\begin{eqnarray}
\Cinv\dCdi &=& \Cinv\dCdi\Cinv\C\\
&=& \sum_k\Cinv\dCdi\Cinv\left(\dCdk\lambda_k+{\bf N}\right)\\
&=& 2 \sum_k F_{ik} \lambda^{(0)}_k+\Cinv\dCdi\Cinv{\bf N}
\end{eqnarray}
where we use the linear dependence of $\C$ on the parameters.
Substituting into \ref{eq:a}, the product of the Fisher matrix with
its inverse leads to a cancellation of the $\lambda^{(0)}$ terms.  We
are left with the final estimator in quadratic form,
\begin{equation}
\label{eq:f}
\hat{\lambda}_i=\frac{1}{2}\sum_j F_{ij}^{-1} \left[\x^T\C^{-1}\dCdi\C^{-1}\x-\Tr\left(\Cinv\dCdi\Cinv{\bf N}\right)\right]
\end{equation}



\twocolumn

\bibliographystyle{mn2e_adslinks}
\setlength{\bibhang}{2.0em}
\setlength{\labelwidth}{0.0em}
\bibliography{refs}

\label{lastpage}

\end{document}